# Giant enhancement of bacterial upstream swimming in macromolecular flows

Ding Cao [1,2,3], Ran Tao [4], Albane Théry [5], Song Liu [6], Arnold J. T. M. Mathijssen [4], and Yilin Wu [1*]

[1] *Department of Physics and Shenzhen Research Institute, The Chinese University of Hong Kong, Shatin, NT, Hong Kong, P.R. China.*

[2] *GBA Branch of Aerospace Information Research Institute, Chinese Academy of Sciences, Guangzhou 510700, China.*

[3] *Guangdong Provincial Key Laboratory of Terahertz Quantum Electromagnetics, Guangzhou 510700, China.*

[4] *Department of Physics & Astronomy, University of Pennsylvania, Philadelphia, PA 19104, USA*

[5] *Department of Mathematics, University of Pennsylvania, Philadelphia, PA 19104, USA*

[6] *Department of Physics and Center for Complex Flows and Soft Matter Research, Southern University of Science and Technology, Shenzhen 518055, P.R. China.*

*To whom correspondence should be addressed.    Mailing address: Room 306, Science Centre, Department of Physics, The Chinese University of Hong Kong, Shatin, NT, Hong Kong, P.R. China.    Tel: (852) 39436354.    Fax: (852) 26035204.    Email: ylwu@cuhk.edu.hk

**Abstract**

Many bacteria live in natural and clinical environments with abundant macromolecular polymers. Macromolecular fluids commonly display viscoelasticity and non-Newtonian rheological behavior; it is unclear how these complex-fluid properties affect bacterial transport in flows. Here we combine high-resolution microscopy and numerical simulations to study bacterial response to shear flows of various macromolecular fluids. In stark contrast to the case in Newtonian shear flows, we found that flagellated bacteria in macromolecular flows display a giant capacity of upstream swimming (a behavior resembling fish swimming against current) near solid surfaces: The cells can counteract flow washing at shear rates up to ~65 $s^{-1}$, one order of magnitude higher than the limit for cells swimming in Newtonian flows. The significant enhancement of upstream swimming depends on two characteristic complex-fluid properties, namely viscoelasticity and shear-thinning viscosity; meanwhile, increasing the viscosity with a Newtonian polymer can prevent upstream motion. By visualizing flagellar bundles and modeling bacterial swimming in complex fluids, we explain the phenomenon as primarily arising from the augmentation of a "weathervane effect" in macromolecular flows due to the presence of a viscoelastic lift force and a shear-thinning induced azimuthal torque promoting the alignment of bacteria against the flow direction. Our findings shed light on bacterial transport and surface colonization in macromolecular environments, and may inform the design of artificial helical microswimmers for biomedical applications in physiological conditions.

**Introduction**

Microorganisms often encounter shear flows in natural and clinical environments where macromolecules are abundant, such as stream biofilms (1), animal digestive tracts (2), and the interstitial space of tissues (3, 4). The response of bacterial motion to shear flows (i.e., rheotaxis) in the presence of macromolecules is therefore critical to diverse microbial processes of environmental and clinical importance, including biofilm formation (5, 6) and pathogen invasion (7). However, bacterial rheotaxis was primarily studied in Newtonian fluids (8-10). Macromolecular fluids commonly display viscoelasticity and non-Newtonian rheological properties (11) that profoundly change bacterial motile behavior in quiescent fluids (12-19). It is much less understood how these complex-fluid properties affect bacterial response to shear flows, either externally applied (20) or self-generated due to collective motion (21).

Here we combine single-cell tracking, flagellar imaging and numerical simulations to study bacterial rheotaxis in various macromolecular fluids. We found that flagellated bacteria in macromolecular flows display a striking capacity of swimming upstream against the flow near solid surfaces: Cells can counteract flow washing at shear rates as high as ~65 $s^{-1}$, and the upstream swimming speed can be up to ~80% of the normal swimming speed; simultaneous imaging of the cell body and flagellar bundle revealed that the cells dominantly steer upstream even at minute shear rates. This behavior represents giant enhancement of upstream swimming compared to the case reported for flagellated bacteria in Newtonian shear flows (9, 22, 23). The phenomenon depends on viscoelasticity and shear-thinning viscosity of the fluid; we explain it as primarily arising from an augmented "weathervane effect" in macromolecular flows due to the presence of a viscoelastic lift force (19) and a shear-thinning induced azimuthal torque promoting the alignment of bacteria against the flow direction. As solid surfaces are prevalent in the living environment of bacteria (6), our findings provide new insight into bacterial transport and surface colonization. The results are also relevant to the near-surface transport of non-flagellated microswimmers (such as spermatozoa (24-26)) and helical microrobots (27) in physiological conditions.

## Results

### Flagellated bacteria display enhanced upstream swimming in macromolecular flows

Motile bacteria powered by rotating flagella display an array of purely physical rheotaxis behaviors (8-10, 22, 23, 28, 29). To study the effect of macromolecules on bacterial rheotaxis near solid surfaces, which is pertinent to the surface lifestyle of bacteria (5), we subjected a flagellated bacterium *Escherichia coli* to shear flows of salmon testes DNA (hereafter referred to as DNA; Methods); DNA solutions are both viscoelastic and shear-thinning (19, 21). Combing precise control of microfluidics and 3D defocused fluorescence microscopy (19, 30) (Fig. 1a), we were able to track bacterial motion near the solid surface (Fig. 1a). To follow cells continuously over sufficiently long periods, we used a smooth-swimming mutant of *E. coli* that does not tumble (HCB1736; Methods).

Remarkably, we discovered that *E. coli* cells in DNA solutions display a giant capacity of swimming upstream against flow, being able to counteract flows and move upstream at near-surface shear rates as high as ~65 s$^{-1}$ (Fig. 1b,c,d; Video 1). Denoting the critical shear rate for bacteria to start being advected downstream as $\dot{\gamma}_c$ (i.e., the shear rate at which the upstream swimming speed becomes zero in Fig. 1c), we found that $\dot{\gamma}_c$ depends on DNA concentration in a non-monotonic manner (Fig. 1d); there exists an optimal DNA concentration (~1700 ng/µL DNA; also see red line in Fig. 1c) at which cells display upstream swimming behavior over the widest range of shear rate up to ~65 s$^{-1}$. Moreover, denoting the swimming speed in quiescent polymer solutions as $V_0$, the maximum upstream swimming speed across different shear rates also depends on DNA concentration non-monotonically and it can be as large as ~80% of $V_0$ (i.e., ~20 µm/s) (Fig. 1e; Fig. S1a; Methods). The highest shear rate that cells can withstand in DNA solutions (~65 s$^{-1}$) is one order of magnitude larger than that reported in Newtonian fluids (~6.4 s$^{-1}$) (23); also the maximum upstream swimming speed (~0.8$V_0$) in DNA solutions is ~4-fold higher than that in Newtonian fluids ~0.18$V_0$ (23) (Fig. S2; Methods). Taken together, our results show that the presence of macromolecules dramatically enhances bacterial upstream swimming.

Meanwhile, along the direction perpendicular or lateral to flow direction, cells always swim to the left of flow direction (i.e., swimming along the positive y-axis direction in the coordinate system specified in Fig. 1a) due to a bias introduced by flagellar chirality (left-handed) (28). The apparent lateral swimming speed in pure growth medium (i.e., a Newtonian fluid) increases monotonically with increasing shear rate (blue line in Fig. 1f), which is consistent with earlier measurement (23, 31). By contrast, the lateral motion bias of *E. coli* cells in DNA solutions depends on shear rate non-monotonically (red and orange lines in Fig. 1f); the maximum lateral motion bias (normalized by $V_0$) across different shear rates decreased with increasing DNA concentration monotonically (Fig. 1g; Fig. 1b; Methods), with excessive DNA concentrations abolishing the lateral bias.   The presence of macromolecules changes the rheotactic swimming pattern of flagellated bacteria both parallel and lateral to the flow directions near a solid surface.

**Two complex fluid properties (viscoelasticity and shear-thinning) are essential to the enhanced upstream swimming**

Enhanced upstream swimming was observed in macromolecular solutions other than DNA (Fig. 2a; Fig. S3), including sodium hyaluronate (HA; Video 2) and polyacrylamide (PAA; Video 3), which indicates that enhanced upstream swimming is a general phenomenon of flagellated bacteria swimming near solid surfaces in macromolecular polymeric flows.   Macromolecular solutions often have viscoelasticity and shear-thinning viscosity (11).   Indeed, the polymer solutions used in our study are viscoelastic with a relaxation time on the order of ~0.01-0.1 s and they display shear-thinning viscosity over the shear-rate range for observing the upstream swimming behavior (Fig. S4) (19) (Methods).

The two rheological properties (i.e., viscoelasticity and shear-thinning) are inter-dependent in many polymer solutions and particularly in our experiments.   To understand the effect of the two complex-fluid properties, we defined a combinatory dimensionless number $De/n$, where $De$ is the dimensionless Deborah number (32) (product of the rotation frequency of flagella and the relaxation time of the polymeric fluid; Table S1; Methods) and $n$ is the flow behavior index (33) ($n < 1$ for shear-thinning fluids; Table S1; Methods).   We plotted the cells' velocity component parallel to flow direction in the plane of shear rate and $De/n$ (Fig. 2b), which can be taken as a 'phase map' of upstream and downstream swimming behaviors.   As shown in the phase map, the range of shear rate supporting upstream swimming behavior (dashed line) strongly depends on $De/n$ and may vary by more than one order of magnitude.   This result shows that the complex-fluid properties of polymer solutions play an important role for upstream swimming of bacteria.   Notably, cells display upstream swimming in a wide range of $De/n$ over several orders of magnitude, indicating that bacteria can swim towards and colonize upstream regions in diverse complex fluid environments.

To account for the effect of viscosity, we examined whether the magnitude of viscosity affects the upstream swimming behavior in polymeric flows.   We tuned the viscosity of macromolecular solutions by adding a viscous agent polyethylene glycol (PEG), which increases viscosity but does not affect either the shear-thinning behavior or the viscoelastic relaxation time (Fig. S5).   We found that, at a given DNA concentration, both the critical shear rate $\dot{\gamma}_c$ (Fig. 2c) for the transition from upstream- to downstream-swimming and the maximum upstream swimming speed (Fig. 2d) decreased with increasing viscosity.   Therefore, a relatively low viscosity is necessary for the upstream swimming behavior.   This result is consistent with the fact that upstream swimming behavior in HA and PAA solutions are not as prominent as that in DNA solutions (Fig. S3), because HA and PAA solutions have higher viscosities at concentrations with comparable normalized upstream swimming speed (Fig. S3, Fig. S4). Interestingly, we note that the maximum lateral motion bias also decreased with increasing viscosity, but the shear rate corresponding to the maximum remained more or less the same at ~5 s$^{-1}$ (Fig. 2e).   In addition, we examined the effect of purely Newtonian polymer solutions on upstream swimming using polymer solutions with negligible relaxation time and negligible shear-thinning behavior, including PEG and Ficoll 400.   Upstream swimming was hardly seen in the shear flow of these Newtonian polymer solutions (Fig. S6; Video 4), and the shear-rate dependence of both velocity components (parallel and lateral to flow direction) was similar to that in polymer-free Newtonian flows (i.e., corresponding to results with 0 ng/µL DNA in Fig. 1c,f), except some anomaly at relatively high concentration of Ficoll 400.

**Flagellar labeling reveals a dominant upstream propulsion behavior in macromolecular flows**

The Eulerian velocity (i.e., the apparent velocity measured in the laboratory frame) of cells analyzed above is a combined effect of flagellar propulsion and passive flow advection.   To delineate the two effects, we developed a high-throughput method to track the velocity direction and the self-propulsion direction of cells by simultaneous imaging of cell body and flagellar bundle in fluorescence microscopy (Fig. 3a; Video 5; Methods).   The self-propulsion direction is determined by the direction from flagellar bundle to the cell body, which is measured by the angle $\varphi$ as shown in Fig. 3a.   The apparent velocity direction of a cell is measured by the angle $\varphi_v$ between upstream direction and the cell velocity component projected on the substrate.   In Newtonian flow with low shear, cells swam in circles with a relatively uniform distribution of $\varphi$ and $\varphi_v$ (Fig. 3b).   As expected, cells swimming in DNA solutions at low shear rates (less than $\dot{\gamma}_c$) had both $\varphi$ and $\varphi_v$ in the range of ~0° to ~90°, i.e., most of them displayed a behavior of upstream propulsion (with lateral bias along the positive y-axis direction in the coordinate system specified in Fig. 1a) and were able to move upstream by counteracting the downstream passive advection (Fig. 3c,d; Fig. S7a).   Unexpectedly, above the critical shear rate $\dot{\gamma}_c$, the self-propulsion direction of cells $\varphi$ remained in the range of ~0° to ~90° while $\varphi_v$ became pointing downstream (Fig. 3e, Fig. S7b), i.e., cells still displayed upstream propulsion but they were not able to counteract the downstream passive advection.   Notably, the apparent moving direction and the self-propulsion direction tended to align with each other (i.e., $|\varphi - \varphi_v|$ near 0° or 180°) as DNA concentration was increased at a given flow rate (Fig. S8).

We further characterized the upstream propulsion behavior by defining an upstream propulsion index $\chi = \langle \cos\varphi \rangle$, where the angular brackets indicate ensemble average over the cell population. An upstream propulsion index near 1 indicates that the mean propulsive force of cells points perfectly upstream. In Newtonian shear flows, we found that $\chi$ was near zero at low shear rate (~<4 $s^{-1}$) and it began to rise as the shear rate was increased (Fig. 3f, blue), suggesting the existence of a critical shear rate for the onset of upstream propulsion. In the presence of DNA polymers, however, the threshold shear rate for the increase of $\chi$ (i.e., the onset of upstream propulsion) appeared to be absent. In particular, even a minute shear rate as low as 1 $s^{-1}$ can trigger strong upstream propulsion with $\chi > \sim0.6$ at relatively high DNA concentrations (>~1000 ng/μL) (Fig. 3f, purple and green). Taken together, these results show a dominant upstream propulsion behavior in macromolecular shear flows, which must be key to the enhanced upstream swimming.

**Model with augmented weathervane effect explains the enhanced upstream swimming in macromolecular flows**

To explain the enhanced ability of bacteria to swim upstream in complex fluids, we build on previously established models for upstream swimming in Newtonian fluids (9, 10, 20, 22, 26, 34, 35). This current literature describes that upstream swimming can be facilitated by a "weathervane effect" (22), which works as follows: The nose of a flagellated bacterium tends to point slightly towards the surface. Consequently, the cell body experiences more friction with the surface than the flagellar bundle, which provides a pivot point. Moreover, the flagellar bundle point slightly away from the surface, with a pitch angle. This pitch exposes the flagella to the flow more than the body. Hence, the bacterium rotates about the pivot point to face upstream like a weathervane in the wind.

Besides the terms considered for Newtonian flows, following ref. (9), we include the contributions of viscoelasticity, shear thinning, and modified swimming speeds (Fig 4; Methods). First, we include the effects of a recently discovered viscoelastic lift force acting on bacterial flagella near solid surfaces (19). This lift force arises from non-linear polymeric stresses and acts in the direction perpendicular to the surface (19). It primarily lifts the flagella but not the cell body due to the comparatively rapid flagellar rotation (Fig. 4a, upper). This leads to a torque that increases the bacterial pitch angle (Fig. 4b), enhancing the weathervane effect. Second, we include another recent finding that the curvature of bacteria swimming in circles near surfaces in a quiescent fluid (36, 37) reduces in polymer fluids (19) (Fig. S9). This curvature reduction can be explained by a model proposing that shear-thinning leads to an additional azimuthal torque that originates from the interactions between the polymers, the bacterium, and the solid surface (38) (Methods). This torque counteracts circular swimming, in favor of upstream alignment (Fig. 4a, lower).   Third, we include the well-known observation that bacterial swimming speeds differ in polymeric fluids (12-15, 39, 40) (for recent reviews, see (41, 42)) (Fig. S9).   A higher self-propulsion speed would allow an upstream-pointing cell to counteract faster counterflows, thus enhancing upstream swimming.   See Methods for a full description of the model.

We combine these effects into a set of coupled stochastic equations of motion, which we integrate numerically for different flow strengths and DNA concentrations (Methods).   This model accurately captures all our findings, specifically that upstream swimming is enhanced in polymer fluids: The model quantitatively predicts both in-plane swimming velocity components as a function of the shear rate (Fig. 1c,f; solid lines) for different DNA concentrations.   Furthermore, it gives an accurate prediction for the critical shear rate (below which upstream swimming is achieved) as a function of DNA concentration (Fig. 1d; solid line), as well as the maximum of the velocity components across different shear rates (Fig. 1e,g; solid lines).   Finally, the model also captures the correct orientation angle distributions for different shear rates and DNA concentrations (Fig. 3b-e; solid lines).   All three different mechanisms (viscoelastic lift force, shear thinning, and speed enhancement) work synergistically to enhance upstream swimming across a broad range of shear rates (Fig. 4c; Methods).

Besides predicting the rheotaxis accurately, our model provides more insight into the dynamics.   We recognize that the viscoelastic lift force can enhance upstream swimming in two ways: On the one hand, the lift force enhances the weathervane effect by raising the flagellar bundle from the surface (Fig. 4a).   This agrees with our measurements for the upstream propulsion index showing that the bacterial polarity is more aligned against the flow as DNA concentration increases at a give shear rate (Fig. 3f).   On the other hand, the lift force points the swimming direction more towards the surface, increasing the tilt angle $\theta$ (Fig. 4b).   The increased tilt angle can reduce the average distance $h$ between the bacteria and the surface so that the bacteria experience less downstream advection.   This result agrees with our experiments showing that $\partial V_x / \partial \dot{\gamma} \sim h$ is smaller for the non-Newtonian fluids at large shear rates, as shown by the less steep slope in Fig. 1c when plotted with linear-scale axes (Fig. S10).   Moreover, the lift force can explain why upstream swimming is reduced at high DNA concentrations (Fig. 1e), because lifting the flagella too high reduces the component of the velocity vector against the flow direction and increases the susceptibility to downstream advection.   Thus, the ability of bacteria to swim upstream is optimized at intermediate polymer concentrations.

**Discussion**

In summary, we have found that two common rheological properties of macromolecular solutions, namely viscoelasticity and shear-thinning, augment the shear-induced weathervane effect experienced by flagellated bacteria and enable giant enhancement of upstream swimming near solid surfaces.  The diagram in Fig. 4d summarizes the mechanisms responsible for the enhanced upstream swimming.  The results shed light on bacterial transport and surface colonization in natural and clinical environments rich in macromolecules.  For example, commensal bacteria dwelling in the digestive tract of animals experience fluid flows consisting of mucus and food-derived macromolecular polymers (2); and pathogens exploring the interstitial space of tissues have to cope with body fluid flows that are rich in biopolymers including collagens and glycosaminoglycans (3, 4).  During these biological processes, the enhanced upstream swimming can rectify the random motion of bacteria and help them collectively explore virgin territories.

Our findings may provide insight for controlling the motion and transport of artificial helical microswimmers (27) under shear flows of polymeric fluids, thus informing the design of microrobotics for biomedical applications in physiological conditions (43, 44).  The study is also relevant to the near-surface transport of non-flagellated microswimmers in polymeric fluids, such as spermatozoa migrating in female reproductive tract. Mammalian spermatozoa display upstream swimming in flows of mucus-rich oviductal fluids (24), and it is intriguing how the non-Newtonian rheological properties of oviductal fluids may impact this motion pattern.  For instance, the steeper near-surface flow profile in the reproductive tract due to shear-thinning viscosity may augment the weathervane effect, enhancing sperm motion against flows and contributing to the fertilization process.

**Methods**

**Experiments**

The following bacterial strains were used: *E. coli* HCB1736 (a derivative of *E. coli* AW405 with *cheY* deletion and smooth-swimming behavior; a gift from Howard Berg, Harvard University, Cambridge, MA) and HCB1737 (a derivative of *E. coli* AW405 and wildtype behavior; a gift from Howard Berg, Harvard University, Cambridge, MA); *E. coli* YW268 and YW191 (*E. coli* HCB1736/1737 transformed with pAM06-tet plasmid carrying kanamycin resistance and expressing green fluorescent protein (GFP) constitutively; the pAM06-tet plasmid was a gift from Arnab Mukherjee and Charles M. Schroeder, University of Illinois at Urbana-Champaign (45)).  Plasmids were transformed via electroporation.  The following polymers were used: salmon testes DNA (abbreviated as DNA; ~2000 base pair, molecular weight $M_w$ ~1.3 MDa, Sigma-Aldrich, cat. No. D1626), polyethylene glycol (PEG, $M_w$ ~8 kDa, Sigma-Aldrich, cat. No. 89510), polyvinylpyrrolidone K90 (PVP, $M_W$ ~0.36 MDa, Sigma-Aldrich, cat. No. 81440), Ficoll 400 ($M_W$ ~0.4 MDa, Sigma-Aldrich, cat. No. F9378), hydroxypropyl methyl cellulose (MC, $M_W$ ~86 kDa, Sigma-Aldrich, cat. No. H7509), sodium hyaluronate (HA, $M_W$ ~1.6 MDa, Hangzhou Singclean Medical Products Co., Ltd., cat. No. 3006700000) and polyacrylamide (PAA, $M_w$ ~5.5 MDa, Sigma-Aldrich, cat. No. 92560).  Similar behavior (enhanced upstream swimming in macromolecular flows) were also found in wildtype bacteria *E. coli* with 'run-and-tumble' motion (Fig. S11).  Although the tumble event is suppressed by solid surface (46) and external flow (47), tumble events still occasionally happened.  The average speed of wildtype cells is less than that of the smooth-swimming mutant (see captions of Fig. S1 and Fig. S11).  Note that tumble events lead to re-orientation and detachment from surface (especially under low flow), and thus make the number of trackable cells less.  A full description of experimental methods and data analysis can be found in Supplementary Materials.

**Numerical model**

***Model summary***.  We build on previously established models for upstream swimming in Newtonian flows, following ref. (9), by including the effects of shear-thinning and viscoelasticity.  Briefly, in Newtonian flows, besides the weathervane effect (9, 22), the rod-shaped cell experiences alignment interaction with the surface due to hydrodynamic and steric effects (48-50), a torque (perpendicular to the surface) that leads to circular swimming in quiescent fluids (36, 37), a torque that leads to Jeffery orbits (51, 52), a torque that arises from the chirality of the flagella (28, 53), and rotational diffusion (54). The sum of all these terms controls the angular dynamics.  Together with self-propulsion and advection, these angular dynamics feed into the translational dynamics, leading to upstream motion.  For macromolecular fluids, we first add the well-known effect that the bacterial swimming speed is modified in polymeric fluids (12-15, 39, 40) (for recent reviews, see (41, 42)). Second, we include a recently proposed effect that shear-thinning leads to an additional azimuthal torque (Fig. 4a) due to coupling between rotation and lateral translation (38, 55). In a quiescent fluid, this effect reduces the curvature of bacteria swimming in circles near a surface (19, 38). To account for the first two effects, we measure bacterial swimming speed and curvature in quiescent fluids with different DNA concentrations (Fig. S9), and import these values directly into our model. Third, for viscoelastic media, we include the recent finding that flagella rotating rapidly near a solid surface experience a viscoelastic lift force (19). This lift force arises from non-linear polymeric stresses and acts in the direction perpendicular to the surface (56), which leads to a torque that increases the tilt angle $\theta$ of the bacteria (see diagram in Fig. 4a). Finally, we combine these effects into a set of coupled stochastic equations of motion, which we integrate numerically for a range of different flow strengths and DNA concentrations. To test this model, we compare the results directly with our experiments.

Specifically, we consider a bacterium composed of an elongated cell body and a left-handed flagellar bundle under shear flow near a single planar surface, far from corners. Our model explicitly takes into account both the in-plane self-propulsion angle $\varphi$ ranging from $-\pi$ to $\pi$ (Fig. 3a), and the tilt angle $\theta$ varying between $-\pi/2$ and $\pi/2$ (Fig. 4a). When $\varphi = 0$, bacteria are oriented upstream, and positive $\varphi$ values denote a drift to the +y-direction, which is the positive vorticity direction. When $\theta = 0$, the bacterium orient parallel to the surface, and positive $\theta$ values correspond to orientations where the cell body points down toward the surface. The bacterium is swimming at a fixed height above the bottom surface, $h$. We assume that $h$ is independent of the strength of the flow, but it depends on the polymer concentration because of the viscoelastic lift force. Close to the wall, the flow profile is linear with a constant shear rate $\dot{\gamma}$ so that bacteria are advected downstream at a speed $V_f = \dot{\gamma} h$. Since both the external flow and the self-propulsion contribute to the dynamics, the bacterial velocity along the *x* and *y* directions can be written as

$$V_x = \dot{\gamma} h - V_0 \cos\phi \cos\theta,$$
$$V_y = V_0 \sin\phi \cos\theta. \qquad (1)$$

The swimming dynamics of the bacteria are therefore fully determined by the flow, the swimming speed, and the tilt and in-plane angles, $\theta$ and $\varphi$, respectively.

***Bacterial orientation dynamics.*** The time evolution of the bacterial orientation is controlled by six contributions. The first four already exist for Newtonian dynamics (9), namely (i) surface alignment $\Omega_\theta^{SA}$, (ii) circular swimming at the wall $\Omega_\varphi^C$, (iii) the weathervane effect, $\Omega_\varphi^{WV}$ and $\Omega_\theta^{WV}$, and (iv) rotational diffusion $D_r$. Additionally, the non-Newtonian nature of the fluid gives rise to (v) an azimuthal torque due to shear thinning, $\Omega_\varphi^{ST}$, and (vi) a viscoelastic lift $\Omega_\theta^{VE}$. Combining these contributions, we have our rheotaxis model as

$$\begin{aligned}\Omega_\theta &= \Omega_\theta^{SA} + \Omega_\theta^{WV} + \Omega_\theta^{VE}, \\ \Omega_\varphi &= \Omega_\varphi^{C} + \Omega_\varphi^{WV} + \Omega_\varphi^{ST}\end{aligned} \quad (2)$$

The bacterial surface alignment is modeled as $\Omega_\theta^{SA}(\theta) = w_{SA} \sin 2(\theta - \theta_0)$, where $w_{SA}$ is an effective angular velocity that accounts for both steric and hydrodynamic interactions (9). The bacteria point slightly towards the wall, with a tilt angle $\theta_0 \sim 10°$ in the absence of flow (54). Another wall effect arises from the counter-rotation of the bacterial cell body and the flagellar bundle near solid surfaces (48-50). The resulting hydrodynamic torque leads to circular motion with a reorientation rate in the $\varphi$ direction, which is approximated by $\Omega_\varphi^C = w_C$. Bacteria with left-handed flagella have positive $w_C$ and exhibit clockwise circling trajectories, as seen from above the surface.

Under shear flow, swimmers near a surface tend to reorient towards the upstream direction due to the weathervane effect (22). The bacterial cell body experiences more friction with the surface than the flagella, which effectively anchors it to the surface so that the flagella are advected down with the flow. Consequently, the weathervane effect realigns bacteria upstream (9, 22, 24-26). We model the resulting reorientation rate as

$$\begin{aligned}\Omega_\theta^{WV}(\theta, \varphi) &= \dot\gamma w_{WV} \cos\varphi \sin\theta^2 \mathbf{1}_{\theta>0} \text{ and} \\ \Omega_\varphi^{WV}(\theta, \varphi) &= -\dot\gamma w_{WV} \sin\varphi \tan\theta \, \mathbf{1}_{\theta>0}\end{aligned} \quad (3)$$

, where $\mathbf{1}_{\theta>0}$ is the Heaviside step function. When the swimmer is oriented away from the surface, the Heaviside step function $\mathbf{1}_{\theta>0}$ considers the reduction in the asymmetry of friction that leads to the disappearance of the weathervane effect.

For the complex fluids, we first account for the fact that the bacterial swimming speed changes (12-15, 39, 40). We measure the intrinsic bacterial swimming speed as a function of the polymer concentration in the absence of flows (Fig. S9a), and directly import these experimental values into the simulations for each concentration. Second, we include the recent finding that the curvature of bacteria swimming in circles near surfaces in a quiescent fluid (36, 37) reduces in polymer fluids (19). This curvature reduction can be explained by a recently proposed effect that shear-thinning leads to an additional azimuthal torque that originates from the interactions between the polymers, the bacterium, and the solid surface (38). The additional azimuthal torque arises from the modified coupling between rotation and lateral translation in the presence of shear thinning (55). To account for this effect, we use a phenomenological expression $\Omega_\varphi^{ST} = w_{ST}$ for the azimuthal torque that counteracts circular swimming ($\Omega_\varphi^C$), in favor of upstream alignment (Fig. 4a).  We determine $w_{ST}$ experimentally by measuring $\kappa$, the curvature of the bacterial trajectories at different DNA concentrations (Fig. S9b).   Note, $\kappa$ is negative for clockwise circles.   Hence, using the previously measured $V_0$ from (Fig. S9a), we compute:

$$w_{ST} = -w_C - \kappa V_0/2\pi, \qquad (4)$$

which is then imported directly into our simulations for each DNA concentration.   In the absence of flows, when $\dot{\gamma}=0$, the measured curvatures from Fig. S9b are recovered in the simulations.   Third, we include the effect of a viscoelastic lift force that raises the rotating flagella up from the surface and into the polymeric fluid (19).   This lift force increases the tilt angle $\theta$ of the cell, thus introducing a new reorientation rate $\Omega_\theta^{VE} = \lambda C_{DNA} \cos\theta$, where $\lambda$ is a constant and $C_{DNA}$ represents the DNA concentration.   In our model for the non-Newtonian fluids, we set this prefactor to $\lambda = 0.001$ and the height is set to $h = 0.5$ µm.   Using these parameters, we find good agreements with the experimental data, including the upstream swimming velocity in Fig. 1 and the bacterial orientation distributions in Fig. 3.

**Stochastic equations of motion**. These orientation dynamics from Eq. (2) are solved numerically using Brownian dynamics simulations with a rotational diffusivity $D_r$.   We initially assign microswimmers with a random in-plane angle $\varphi$ and a zero tilt angle $\theta$. We then integrate the angular dynamics in a non-Euclidian coordinate system (57) using the Langevin equations:

$$\delta\theta = \Omega_\theta \delta t + \tan\theta \, D_r \delta t + \sqrt{2D_r \delta t}\xi_\theta,$$

$$\delta\varphi = \Omega_\varphi \delta t + \frac{\sqrt{2D_r \delta t}}{\cos\theta}\xi_\varphi. \qquad (5)$$

We simulate these stochastic equations for a population of $N = 10^4$ swimmers with a time step of $\delta t = 0.1$ s.   After computing the swimmer orientations, the spatial dynamics are found by integrating Eq. (1).   A simulated trajectory continues until it reaches the simulation time limit, $t_f = 1000$ s, or when the tilt angle exceeds a designated escape angle, $\theta_e$, when the swimmer leaves the surface.

The model parameters are determined from experiments. We use the surface-alignment coefficient $w_{SA} = 4.0$ s$^{-1}$, the weathervane effect coefficient $w_{WV} = 0.075$ s$^{-1}$, the head-tail rotation coefficient $w_C = 0.49$ s$^{-1}$, and height $h = 0.9$ µm to match the experimental data for Newtonian fluids. For the escape angle we use $\theta_e = -20°$, and for the rotational diffusivity we use $D_r = 0.057$ µm$^2$s$^{-1}$   (9, 20, 54, 58, 59).

To quantify the contributions from each of these different mechanisms, i.e., viscoelastic lift force (VE), shear thinning (ST), and speed enhancement (SE), separately and together, we compare variations of our theoretical model with increasing complexity (Fig. 4c).   We first simulate bacterial rheotaxis in a Newtonian fluid (blue line; same as in Fig. 1c).   We then use the same equations of motion but with an increased swimming speed as measured in polymer fluids (SE, dashed line in Fig. 4c): This speed enhancement alone does not result in significant upstream swimming.   Next, we add the curvature reduction resulting from shear-thinning (SE+ST, green dashed line in Fig. 4c): This effect enhances upstream swimming, especially at lower flow speeds; this result agrees with the independent experimental study by Maldonado et al. (ref. (38)) that we came to be aware during the preparation of this manuscript.   Subsequently, we consider the speed enhancement and viscoelasticity only (SE+VE, purple dot-dashed line in Fig. 4c): Here we observe upstream swimming at high flow speeds, but not at low flow speeds, because the viscoelastic effects increase with shear rate.   Finally, we add all effects (SE+ST+VE, red long-dashed line in Fig. 4c): Together, these synergistically add to upstream swimming across a broad range of shear rates, as observed in the experiments (red data points in Fig. 4c).

**Supplementary Materials**, including Supplementary Methods, Supplementary Figures, Supplementary Table and Videos, are available in the online version of the paper.

**Acknowledgements.** We thank Kaiwei Wang for help with HA experiments; Arnab Mukherjee and Charles M. Schroeder (University of Illinois at Urbana-Champaign), and Howard Berg (Harvard University) for generous gifts of bacterial strains. Y.W. and D.C. acknowledge support by the Ministry of Science and Technology Most China (No. 2020YFA0910700, Y.W.), National Natural Science Foundation of China (NSFC No. 31971182, Y.W.; NSFC No. 61988102, 62305069, D.C.), the Research Grants Council of Hong Kong SAR (RGC Ref. No. 14306820, 14307821, RFS2021-4S04 and CUHK Direct Grants, Y.W.), the Natural Science Foundation of Guangdong Province of China (2024A1515010367, D.C.), and the Guangzhou Municipal Science and Technology Planning Project (2023A04J0339, D.C.). A.J.T.M.M. acknowledges funding from the United States Department of Agriculture (USDA-NIFA AFRI grants 2020-67017-30776 and 2020-67015-32330), the Charles E. Kaufman Foundation (Early Investigator Research Award KA2022-129523) and the University of Pennsylvania (University Research Foundation Grant and Klein Family Social Justice Award). A.T. received support from the Simons Foundation through the Math+X grant awarded to the University of Pennsylvania. Y.W. acknowledges support from New Cornerstone Science Foundation through the Xplorer Prize.

**Author Contributions**: D.C. discovered the phenomena, designed the study, performed experiments, analyzed and interpreted the data. R.T, A.T. and A.J.T.M.M developed the theory and numerical model and performed simulations. S.L. helped with initial experiments and manuscript preparation. Y.W. conceived the project, designed the study, analyzed and interpreted the data. A.J.T.M.M. supervised theory development and coordinated communications between different groups regarding theory design, data analysis and interpretation of the results. Y.W. wrote the paper with D.C., A.J.T.M.M. and other authors' input.

**Author Information**: Reprints and permissions information is available at the journal website. The authors declare no competing financial interests. Requests for materials should be addressed to Y.W. (ylwu@cuhk.edu.hk).

**Figures**

Figure 1

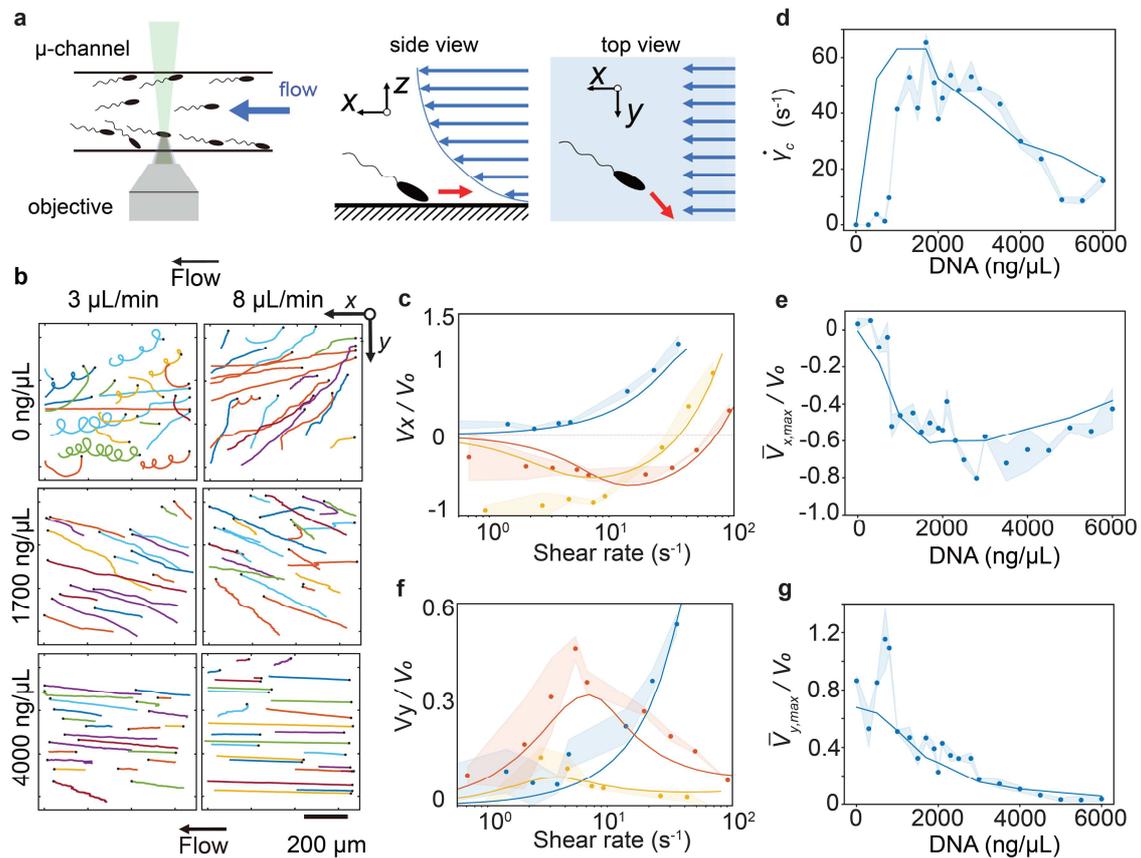

**Fig. 1. Enhanced upstream swimming behavior of *E. coli* in macromolecular shear flows.** (a) Schematic diagram of the experimental setup. Dilute suspensions of *E. coli* (smooth-swimming mutant; YW268) were flowed through a microfluidic channel, with the flow rate controlled by a syringe pump, and cells near the bottom surface were imaged by 3D defocused fluorescence microscopy (left sub-panel; Methods). A bacterium swimming near the bottom surface of the channel (middle: side view, *x-z* plane; right: top view, *x-y* plane) encounters the external shear flow (blue arrows) imposed along the positive *x* direction; red arrow indicates the apparent moving direction of the bacterium in laboratory reference frame. In most cases cells at ~5 μm above the surface were swept downstream by the flow. (b) 2D projection of cells' 3D trajectories tracked near glass surface at different DNA concentrations and flow rates (Methods). Each trajectory starts with a black dot. Scale bars, 200 μm. (c,f) Ensemble average of swimming velocity components parallel ($V_x$) and lateral ($V_y$) to flow direction (normalized by the mean speed of cells in the absence of flow, $V_0$) plotted against near-surface shear rate (Methods). Experimental data (dots) are compared directly with simulation results

(solid lines). Different colors indicate DNA concentration in the fluid: blue, 0 ng/μL; red, 1700 ng/μL; yellow, 4000 ng/μL. Panels c,f share the same $V_0$ for a specific DNA concentration: 27.3 μm/s, 38.8 μm/s and 21.3 μm/s for blue, red and yellow data, respectively. Negative (or positive) $V_x$ corresponds to upstream (or downstream) motion, while positive (or negative) $V_y$ corresponds to lateral motion bias to the left (or right) of flow direction. Dashed lines connecting data points serve as guides to the eye. (d) Critical shear rate $\dot{\gamma}_c$ plotted against DNA concentration. (e) Maximum of $V_x$ (denoted as $\bar{V}_{x,max}$; normalized by $V_0$) plotted against DNA concentration. (g) Maximum of $V_y$ (denoted as $\bar{V}_{y,max}$; normalized by $V_0$) plotted against DNA concentration. The value of $\bar{V}_{x,max}$ (or $\bar{V}_{y,max}$) for a specific DNA concentration was obtained from data similar to those shown in panel c (or panel f) (Methods); $|\bar{V}_{x,max}|$ and $\bar{V}_{y,max}$ correspond to the maximum upstream swimming speed and the maximum lateral motion bias in shear flows at the DNA concentration. Error bars (shaded area) in panels c-g indicate standard error of the mean (N=3 replicate experiments).

Figure 2

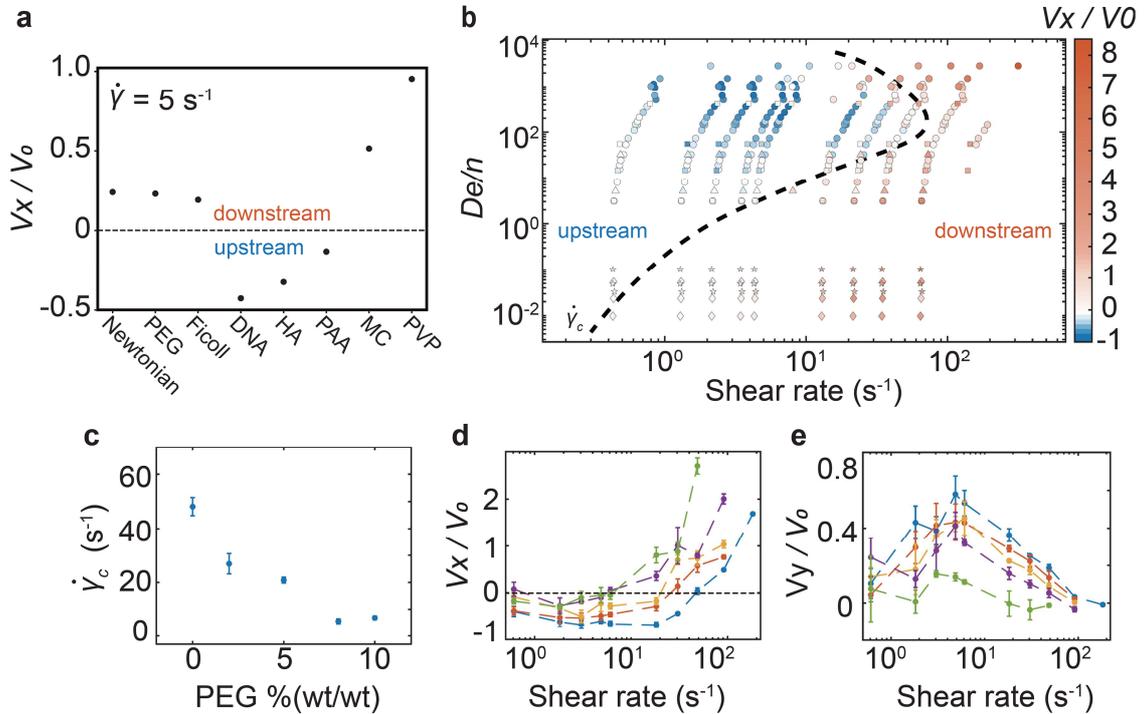

**Fig. 2. Effect of rheological properties on *E. coli* swimming behavior in shear flows of polymer solutions.** (a) Swimming velocity along the flow direction ($V_x/V_0$) against shear rate ($\dot{\gamma}$) in different media: Newtonian, 10% PEG, 5% Ficoll, 1700 DNA, 0.1% HA, 0.5% PAA, 0.3% MC, 3% PVP. The intrinsic swimming speed in these solutions in the absence of flow is $V_0$ = 27.3, 22.3, 30.1, 38.8, 27.9, 28.8, 55.8, 25.6 µm s$^{-1}$, respectively. Error bars indicate standard error of the mean (N = 3 replicate experiments). (b) Swimming velocity component parallel to flow direction ($V_x$) (normalized by the mean speed of cells in quiescent fluids, $V_0$) plotted in the plane of near-surface shear rate and the combinatory dimensionless number $De/n$. Color bar indicates the value of normalized $V_x$, and therefore blue (or red) regions correspond to the regimes of upstream (or downstream) swimming. Data in this figure was acquired with solutions of DNA (circles), HA (squares), PAA (triangles), PVP (stars) and MC (diamonds) (Fig. S4). The dashed line indicates critical shear rate $\dot{\gamma}_c$. (c) Critical shear rate $\dot{\gamma}_c$ at different PEG concentrations with the DNA concentration held constant at 1700 ng/µL. (d,e) Swimming velocity components parallel ($V_x$) and lateral ($V_y$) to flow direction (normalized by $V_0$) plotted against near-surface shear rate in the same manner as Fig. 1c,f. DNA concentration in the fluids was held constant at 1700 ng/µL, and different colors indicate PEG concentration (wt/wt): blue, 0%; red, 2%; yellow, 5%; purple, 8%; green, 10%. The value of $V_0$ for a specific PEG concentration in panels d,e is 29.2 µm/s, 23.7 µm/s, 18.1

µm/s, 15.2 µm/s and 13.3 µm/s for blue, red, yellow, purple and green data, respectively. Error bars in panels c-e indicate standard error of the mean (N=3 replicate experiments).

Figure 3

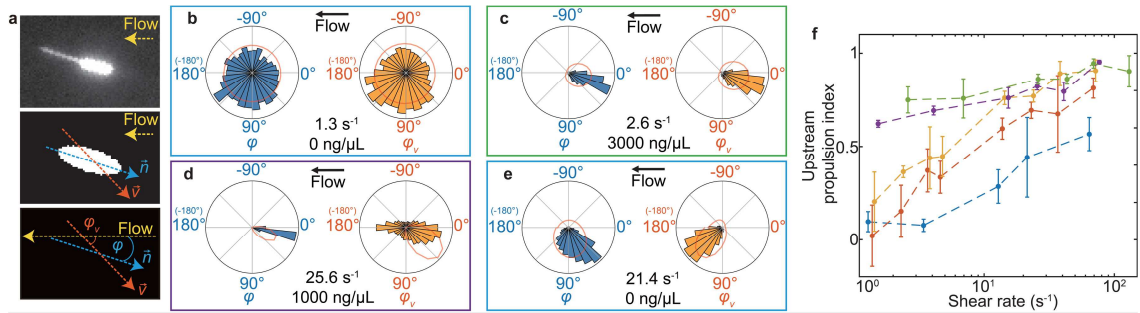

**Fig. 3. Measurement of self-propulsion direction reveals dominant upstream propulsion of *E. coli* in macromolecular shear flows.** (a) Schematic of self-propulsion direction measurement (Methods). Upper: Original grayscale fluorescence image of an *E. coli* cell (cell body and flagella simultaneously labeled and imaged). Middle: The grayscale image is binarized with appropriate intensity threshold and fitted to an ellipse, with the major axis of the ellipse taken as the in-plane self-propulsion direction $\vec{n}$ (pointing from flagella to cell body; blue arrow); the apparent swimming velocity component projected on the substrate $\vec{v}$ (red arrow) is calculated via cell tracking (Methods). Lower: The angle between $\vec{n}$ (or $\vec{v}$) and flow direction (yellow arrow) is defined as $\varphi$ (or $\varphi_v$). Both $\varphi$ and $\varphi_v$ ranges from -180° to 180°. (b-e) Distributions of *E. coli* self-propulsion direction $\varphi$ and apparent swimming velocity direction $\varphi_v$ at shear rates and DNA concentrations indicated in the panels. Near-surface shear rate is indicated in each sub-panel, with the volume flow rate being 0.3 μL/min (panel b,c) and 5 μL/min (panel d,e), respectively. Experiments (histograms) are compared directly with simulation results (red lines). (f) Upstream propulsion index $\chi = \langle \cos \varphi \rangle$ plotted against near-surface shear rate at different DNA concentrations (blue: 0 ng/μL, red: 500 ng/μL, yellow: 800 ng/μL, purple: 1000 ng/μL, and green: 3000 ng/μL). Error bar indicates standard error of the mean (N=3 replicate experiments).

Figure 4

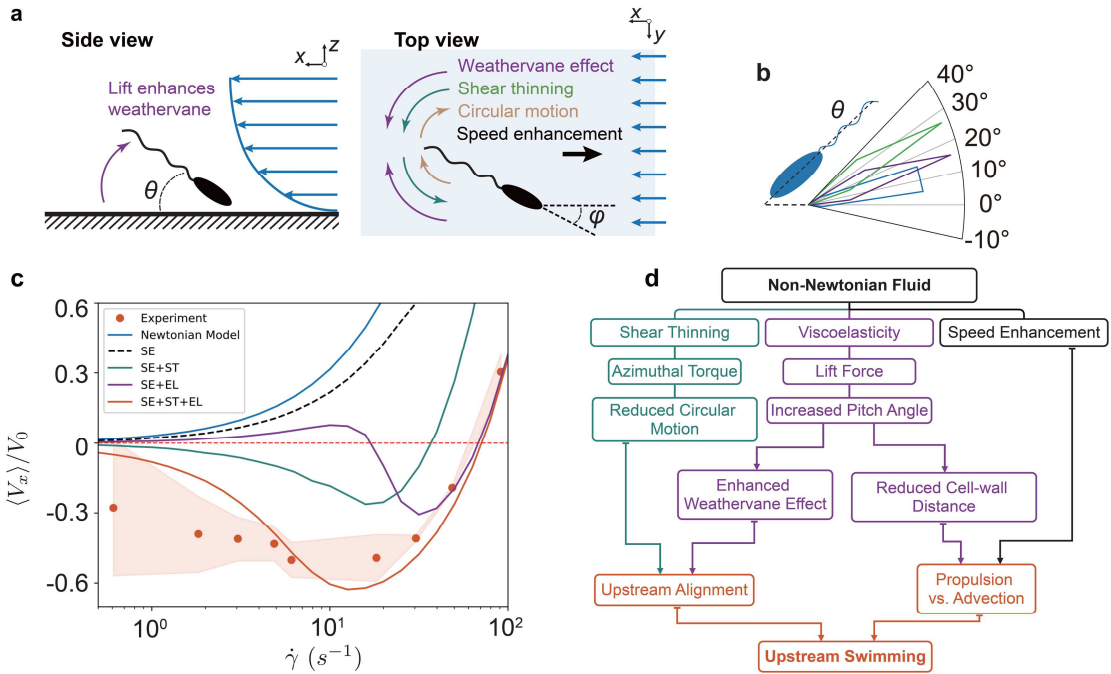

**Fig. 4. Conceptual model of enhanced bacterial upstream swimming in macromolecular shear flows**. (a) Schematics of a bacterium swimming near a solid surface in macromolecular shear flows. Side view: The cell experiences a viscoelastic lift force (VE) primarily acting on the flagella, increasing the pitch angle (purple arrow). Top view: This increased the pitch angle enhances the weathervane effect that reorients the bacteria in the upstream direction. At the same time, shear-thinning (ST) leads to an azimuthal torque that counteracts the torque responsible for swimming in circles, again favouring the weathervane effect. Finally, non-Newtonian effects lead to a swimming speed enhancement (SE) that promotes upstream propulsion over downstream advection. (b) Numerical simulation: distributions of the tilt angle, θ, simulated for three DNA concentrations (blue: 0 ng/μL, purple: 1000 ng/μL, and green: 3000 ng/μL). (c) Comparison between different theoretical models that feature speed enhancement (SE), shear thinning (ST), and/or viscoelastic lift force (VE). These models (lines) are compared to experimental data (markers) for bacteria swimming in a 1700 ng/μL DNA solution. Error bars (shaded regions) indicate standard error of the mean (N=3 replicate experiments). (d) Diagram summarising the mechanisms responsible for enhanced bacterial upstream swimming in Non-Newtonian fluids.

25